# Synthetic fast ion diagnostics in tokamaks: comparing the Monte Carlo test particle code ASCOT against experiments

Simppa Äkäslompolo, Taina Kurki-Suonio, Seppo Sipilä, and the ASCOT Group

*Aalto University, Department of Applied Physics, Finland*

**Corresponding author:**

Simppa Äkäslompolo
Aalto University
Department of Applied Physics
P.O. Box 14100
FI-00076 Aalto, Finland
simppa.akaslompolo@alumni.aalto.fi

**Abstract**

Measuring fast ions, most notably fusion alphas, in ITER and future reactors remains an issue that still lacks an adequate solution. Numerical simulations are invaluable in testing the potential and limitations of various proposed diagnostics. However, the validity of the numerical tools first has to be checked against results from existing tokamaks. In this contribution, a variety of synthetic diagnostics for fast ions (collective Thomson scattering, neutral particle analyzer, neutron camera, infrared measurements, fast ion loss detector and activation probe) from the orbit-following Monte Carlo code ASCOT are compared to measurements from several tokamaks (ASDEX Upgrade, DIII-D and JET). Within the limitations of physics included in the numerical model and availability of input data from experiments, the agreement between synthetic data and measurements is found to be quite good.

**Keywords**

ASCOT, Monte Carlo simulation, fast ion diagnostics, tokamak

# I. INTRODUCTION

Modelling of diagnostics as part of more comprehensive plasma simulations is important not only for the optimization of the experimental setup but, at least in the case of energetic particles (most notably fusion alphas), also for building a more comprehensive picture of the physical reality from the limited data obtained by the diagnostics. The synthetic diagnostics of simulation codes also work the other way around: they provide a way to validate the codes by comparing synthetic data against physical measurements.

ASCOT[1] is a test particle orbit-following Monte Carlo code for toroidal magnetic fusion devices. The code solves the distribution of minority species by following the trajectories of the corresponding test particles. The particles undergo collisions with a static Maxwellian background plasma. The detailed magnetic fields and the first wall can be fully three-dimensional. The code models exactly the neoclassical and classical transport of particles as well as any effects due to a toroidally asymmetric magnetic field, but it also features a model for MHD modes relevant for fast ions (neoclassical tearing modes and toroidal Alfvén eigenmodes).

ASCOT is mainly used for transport studies of fast ions and impurities in realistic tokamak geometries, but the recent addition of a fully relativistic collision operator and the synchrotron radiation force also facilitate simulating runaway electrons. ASCOT has been developed and maintained since the early 1990's. Since then, there have been multiple projects aimed at (or benefitting) code validation and improved interpretation of diagnostic measurements by comparing experiments to ASCOT simulations. The results range from conceptual designs to full quantitative comparisons.

Numerical models can probably never encompass all the physics in the processes they are simulating, but if the dominant ones are properly implemented, the agreement with experiments should be satisfactory. The purpose of this contribution is to probe the validity of the ASCOT code by comparing its predictions to the measurements from not just one but a variety of different fast ion diagnostics that probe either the population of confined fast ions or those lost from the plasma. Much of the data have already been published and are mainly reviewed here in order to present the whole of the various validation efforts. The ASCOT simulation results are compared against measurements from collective Thomson scattering (CTS), neutral particle analyzer (NPA), fast ion loss detector (FILD), activation probe, neutron camera, and infrared camera. These measurements have been carried out at JET, ASDEX Upgrade and DIII-D. For some diagnostics, even quantitative (when possible) agreement between synthetic and physical measurements is quite good (neutron camera, infrared measurements), while for other diagnostics (certain cases with NPA) even qualitative agreement is poor. In the latter case, sources for the discrepancies are identified.

## II. COMPARISONS OF ASCOT SIMULATIONS AGAINST VARIOUS DIAGNOSTICS

### II.A. Collective Thomson Scattering (CTS)

Collective Thomson Scattering (CTS) is one of the few measurement techniques capable of probing the confined fast ion distribution.[2] One-dimensional fast ion velocity distribution functions from CTS experiments at ASDEX Upgrade have been compared with simulations using the codes TRANSP/NUBEAM[3] and ASCOT for two different neutral beam injection (NBI) configurations, one with a single NBI source and the other with two sources. For more details on the experiments themselves, see Ref.4.

The two simulation codes compute the two-dimensional fast ion distribution $f(v_\parallel, v_\perp)$ at the location of the scattering volume, where $v_\parallel$ and $v_\perp$ are the velocity components parallel and perpendicular to the magnetic field, respectively. This distribution was projected onto the CTS measuring direction, i.e. along the fluctuation wave vector $\mathbf{k}_\delta = \mathbf{k}_s - \mathbf{k}_i$, where the subscripts $s$ and $i$ refer to scattered and incident radiation. The projection $g(u)$, where $u$ is the resolved one-dimensional velocity component along the fluctuation vector, was then plotted together with the results of CTS measurements.

The two simulation codes were found to be in good agreement with each other. The CTS measurements are sensitive to a number of things with well-known uncertainties, including the background ion temperature, receiver calibration, beam overlap and probing power. Considering all these uncertainties, the first comparisons of CTS results to numerical plasma simulations yield a reasonable level of agreement, as shown in Figure 2.

**II.B. Neutral Particle Analyser Measurements**

Neutral particle analyzer (NPA) is also capable of extracting information on the confined fast ion populations, albeit only as a line-integrated signal.[5] Comparisons of ASCOT's NPA simulation model to data from the ASDEX Upgrade tokamak have been made and were reported in Ref. 6.

In ASCOT, the line-of-sight of the NPA spans a finite cone and signal is collected according to particle energy to bins corresponding to different energy channels. Whenever a test particle intersects the cone, its velocity components are checked. If the direction of motion allows reaching the detector, the particle's neutralization probability is evaluated and it is assumed to travel ballistically to the detector. Along the way, the attenuation of its signal is calculated from the background plasma density and temperature.

In 2005, several ASDEX Upgrade discharges were dedicated to comparisons of ASCOT simulations to measurements. The NPA sightlines were varied from discharge to discharge. In each discharge six neutral beam injection sources were turned on in sequence. Measurements with three different NPA orientations of the high energy neutral deuterium flux of re-neutralized NBI ions were compared to simulations. Comparisons between measurement and simulation are shown in Figure 3.

Simulations of radially launched 60 keV neutral beams measured with radial NPA lines-of-sight were found to match the measurements well even quantitatively. The 90 keV tangential beams or tilted NPA lines of sights, however, did not produce satisfactory agreement. There are two primary candidates for the source of discrepancy. The background neutral density, playing a crucial role in both the neutralization of the fast ion contributing to the signal and attenuation of the signal, is poorly known experimentally. Therefore it was simply modelled by an exponential function, decaying radially inward from the separatrix. However, in reality, neutral density is (at least) two-dimensional and thus any deviation from the radial direction when comparing measurements to synthetic data is known to contain a large uncertainty. The other candidate for explaining the discrepancy for tangential beams is the observed but yet-to-be-understood anomalous redistribution of fast ions. Tangential beams produce wider orbits at the edge region and, thus, are probably more susceptible to anomalous processes.

**II.C. Neutron Camera Simulations**

In a study of tungsten off-axis accumulation in rotating JET plasmas, 2D tungsten density profiles were used in ASCOT to model the change in neutral beam deposition. The predicted beam-target neutron rates were then compared to those measured by the vertical KN3 neutron camera. For more details on the experiments, see Ref. 7.

Since beam-target reactions provide the dominant deuteron-deuteron (DD) fusion reaction channel in JET deuterium plasmas, the resulting 2.5 MeV neutrons are a good proxy for diagnosing the fast ion distribution. A synthetic neutron camera signal, comparable to the measured signal, was obtained by first calculating the DD reaction rate profile in (R,z) by ASCOT using parametrized fusion cross-sections, and then integrating the profile along the lines of sight of the detector channels, neglecting neutrons born outside the viewing cone that can reach the camera via scattering.

Figure 4 shows the experimental and synthetic signals corresponding to the horizontal and vertical channels of the neutron camera KN3. For the comparison, the experimental signal was averaged over 30ms around the simulated time slices and corrected with geometric factors. The averaging time was chosen to be short enough to capture the effect of off-axis tungsten peaking within one inter-ELM (Edge Localized Mode) period. Two discharges, 82722 at 5.9s and 82794 at 5.3s, were analyzed. The agreement is quite good even quantitatively. In discharge 82722, however, the central vertical channels see roughly 20% less neutrons than predicted by ASCOT. This may be due to a sawtooth crash slightly before 5.9s, not accounted for in the simulation due to the lack of a sawtooth mixing model.

**II.D. Localized Neutral Beam Ion Wall Loads due to TBM Mock-Up Coils**

Infrared measurements provide information about the changes in the temperature of material components emitting the radiation and, thus, of the power loads arriving at the component. Therefore such measurements can be used to give indirect information on changes in plasma confinement due to various perturbations.

The effect on fast ion confinement of the so-called test blanket modules (TBM), to be used in ITER to demonstrate tritium breeding, was simulated in DIII-D using mock-up coils installed near the plasma.[8] The coils produced a magnetic perturbation similar to but significantly stronger than the one produced by the ferromagnetic material in ITER TBMs. During the experiments, the mock-up coils were switched on and off, while the wall power loads were recorded by infrared measurements.

The various beams in DIII-D were injected separately to measure the TBM-induced hot spots for different pitch-angle distributions. With neutral beam injection, the magnetic fields generated by the mock-up coils were shown to cause a hot spot on the two central carbon tiles protecting the mock-up coils. Furthermore, it was found that this hot spot only appears during NBI injection and, therefore, it was concluded that the hot spot is due to fast ion losses.

These findings were corroborated by fast ion loss simulations. The measurements were compared to simulated wall loads from three fast ion simulation codes (ASCOT, OFMC[9] and SPIRAL[10]) that can follow the particles all the way to the DIII-D wall. Figure 5 displays both the infrared measurements and the power loads from the three fast ion codes. The data is shown for five different beam configurations. Calculating the power load distribution in the protective tiles from the infrared measurements is difficult due to complications in heat diffusion, but the peak heat loads (indicated as numbers in Figure 5) are immune to these difficulties. This is because the hottest spot is entirely due to the fast ions hitting the spot and, thus, should be faithfully reproduced by fast ion simulations. Indeed, the peak values match better than the heat flux distributions. From the reasonable agreement between the experimental and synthetic data, particularly for ASCOT and OFMC since they had a more sophisticated first-wall model, it was concluded that these fast-ion codes can be used to estimate TBM-enhanced power loads also in ITER.

**II.E. Fast Ion Loss Detector Measurements of Neutral Beam Ions**

The Fast Ion Loss Detector (FILD) is a scintillator base diagnostic for fast ions leaving the plasma.[11] FILD allows resolving not only the energy of the lost ions but also their pitch. The effect of the in-vessel coils on fast ion wall power loads was simulated and compared to FILD measurements in ASDEX Upgrade.[12] Neutral beam injected ions were simulated in two discharges (#26476 and #26895) both in the presence and in the absence of the magnetic field perturbation induced by the eight newly installed in-vessel coils. In discharge #26476 the beams were applied individually, making it a useful basis for investigating the effect of the coils on different beam orientations and, thus, on fast ions with different pitches.

To achieve maximal realism, ASCOT could in principle model the casing, the collimator, and the scintillator plate of the FILD. However, due to the small size of the collimator slit, limiting the pitch range of the ions incident on the plate would allow only a tiny fraction of the test particles to actually hit the plate, making the statistics unacceptably low. Therefore, the pitch and energy distribution of all particles that hit the casing were statistically analyzed.

Even though the ion optics of the probe spread the details into a wide Gaussian peak,[13] the results from the ASCOT synthetic diagnostic were found to correspond well with the FILD measurements, see Fig. 6. The strong peak at approximately 70°/40mm is at the correct location, but is unfortunately so strong in the experiment that it drowns most of the other features. Furthermore, the experimental signature is seen to extend up to 50mm in Larmor radius while the simulated signal has a sharp edge at 40mm. This is because in the simulations, the signal from test particles is undistorted, while the physical device has an instrument function, generally of Gaussian shape, that smears the signal into a wider range, blurring the peaks. There are further effects from the omitted ion optics: the ions with gyro radius less than roughly 20mm or pitch angle less than 30° cannot be physically measured. This means that the weak peak at around 30° in the synthetic data is barely measurable at all, but is just visible in the absence of the in-vessel coil perturbation.

**II.F. Fusion Product Activation Probe Measurements**

Fast ions escaping the plasma can produce radioactivity in certain materials. This radioactivity can then be measured post-mortem or, ideally, even during the discharge. This is the principle behind, e.g., the fusion proton activation probe.[14] ASCOT simulations of the fusion product flux have been compared to fusion proton activation probe measurements in ASDEX Upgrade.[15]

In the adjoint Monte Carlo integration scheme that was employed, the roles of the relatively large source, i.e., the plasma, and the tiny target, i.e., the target samples within the probe, are reversed. Since the target is small (and only visible through the narrow slit in the graphite shell) and the source is large, only very few markers launched from the plasma will find their way to the target. Therefore, most of the markers do not contribute to the measurement. In the adjoint method, markers start backward in time from the target and are much more likely to pass through the plasma. The adjoint density is closely related to the "instrument function" of the probe: It directly indicates which parts of the plasma the probe measures and with what kind of relative sensitivity.

The calculation of the flux was done in three phases: calculation of the fusion reactivity on an (R,z) grid, calculation of the adjoint density (R,z) grid, and multiplication of these two together. This method was applied to the activation probe experiments at ASDEX Upgrade.[15] The resulting flux calculated for discharge #29226 is presented in Figure 7 for the three different fusion reactants (H, T and $^3$He) and three different means providing the reactions (thermal fusion, beam-target fusion, and beam-beam fusion). The experimentally measured total number of particles with measurement uncertainty was divided by the flat top length of 6.0s to calculate the number of arriving particles per second. The simulated signal has the right order of magnitude, being within a factor of about 2 of the experimental measurement, which can be considered very good considering all the uncertainties in the plasma profiles and the total neglect of effects such as ELMs that, from FILD measurements, are known to affect fast ions. The numerical analysis provided, for the first time, an absolutely calibrated flux of fusion products to the probe.

**II.G. Measurements and Simulations of NBI Fast Ion Wall Loads in the Presence of Toroidal Field Ripple**

In preparation for ripple experiments at JET, power loads on the plasma-facing components due to fast ions from external heating – neutral beam injection (NBI) and ion cyclotron resonance heating (ICRH) – were calculated using appropriate codes.[16] In 2006–2008, for the NBI-induced power loads, orbit-following Monte Carlo codes OFMC and ASCOT were used extensively to prepare for and, later on, to analyze the JET ripple campaigns. ASCOT was also used to do the same for ICRH-generated ions. OFMC and ASCOT have been extensively benchmarked and a JET-relevant wall load benchmark of the codes against each other is presented, e.g., in Ref. 17. Nonetheless, it was seen fit to use two independent codes to verify the loads.

During the experiments, radiation expected to be caused by the losses of NBI ions was measured using visible-light and infrared cameras. Figure 8(a) shows a comparison of the peak power load deduced from the IR measurement and a simulated one, in this case by the OFMC code. The agreement between the predictions from the simulations and the measurements in Fig. 8(a) is very good even with the linear scale used in the figure. Figure 8(b) shows the nearly identical heat load patterns produced by the two codes, with the maximum loads deviating less than 10%. It can thus be inferred that the comparison to IR camera measurements shown in Fig. 8(a) is valid for both codes.

## III. CONCLUSIONS AND DISCUSSION

The validation efforts outlined in this contribution show that test particle codes like ASCOT can successfully be applied to complement experimental measurements by revealing what kind of particle distribution is responsible for the measured signals. The cases addressed in this contribution were all concerned with fast ion distributions. In some cases the agreement with experimental measurements were surprisingly good even quantitatively (activation probe), while in some other cases even the qualitative behavior left a lot to be desired (off-axis beams with tangential NPA sightline). In applying test particle code results for inferring information on the particle distribution responsible for the signal, it is important to carefully check that all the physics present in the experiment is also present in the simulations. Sometimes, for instance with ELMs, this is very difficult if not impossible, and then it is to be expected that some discrepancies will remain when comparing synthetic signals (that correspond to snapshots) to such experimental signals that are obtained by averaging over time.

**Acknowledgments**

This work was partially funded by the Academy of Finland project Nos. 121371, 134924 and 259675, Tekes – the Finnish Funding Agency for Innovation under the FinnFusion Consortium, and the US Department of Energy under DE-AC02-09CH11466, SC-G903402, DE-FC02-04ER54698 and DE-AC05-00OR22725.

This work, supported by the European Communities under the contract of Association between Euratom/Tekes, was carried out within the framework of the European Fusion Development Agreement. The views and opinions expressed herein do not necessarily reflect those of the European Commission or the ITER Organization. The supercomputing resources of the CSC - IT Center for Science were utilized in the studies.

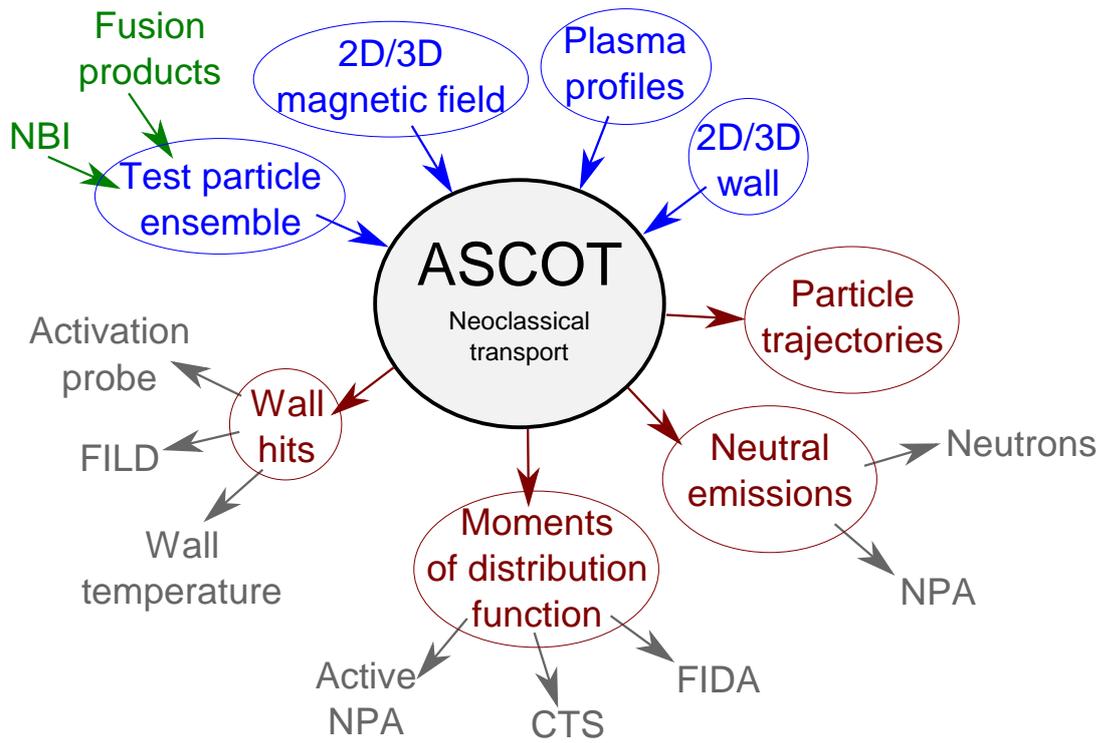

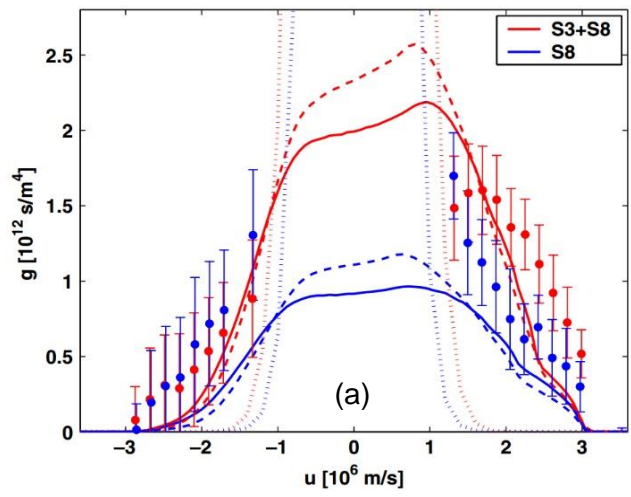 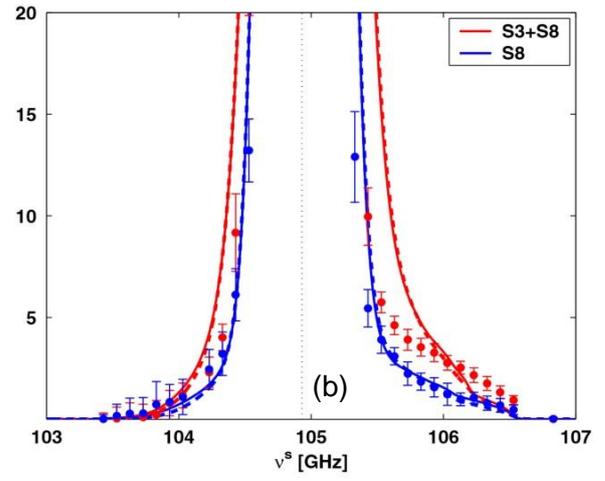

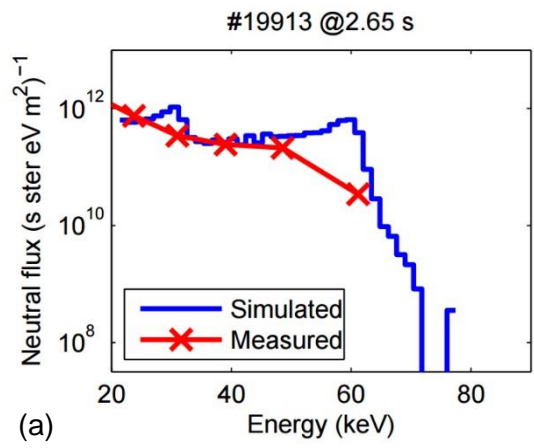 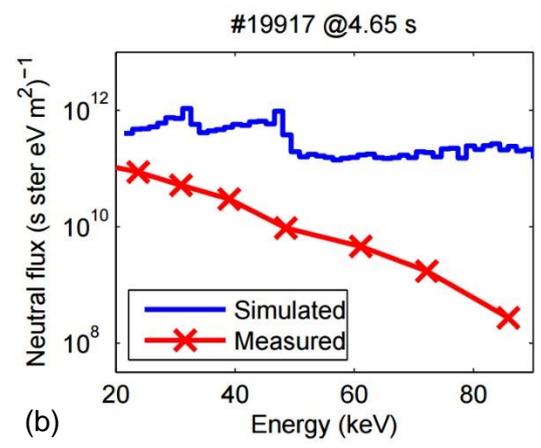

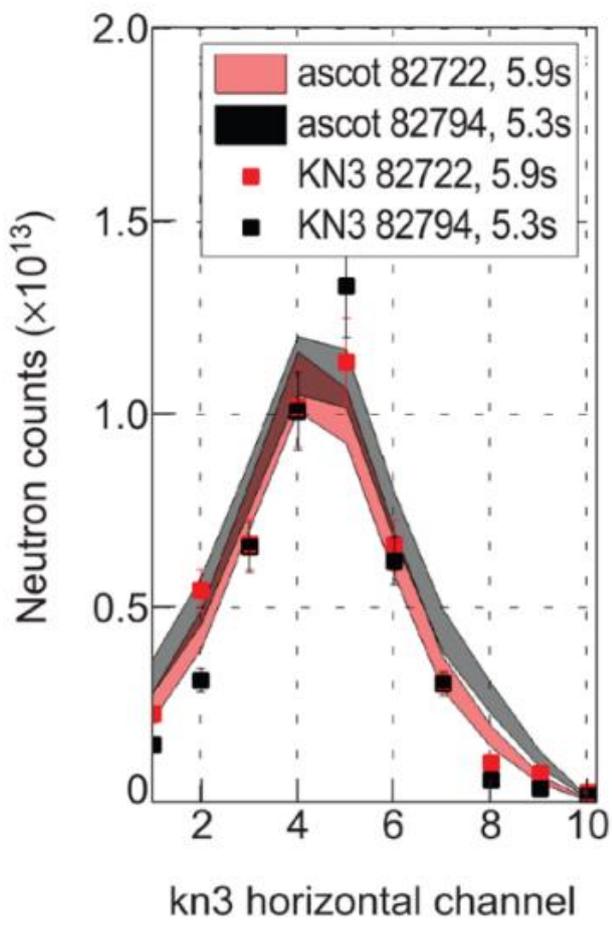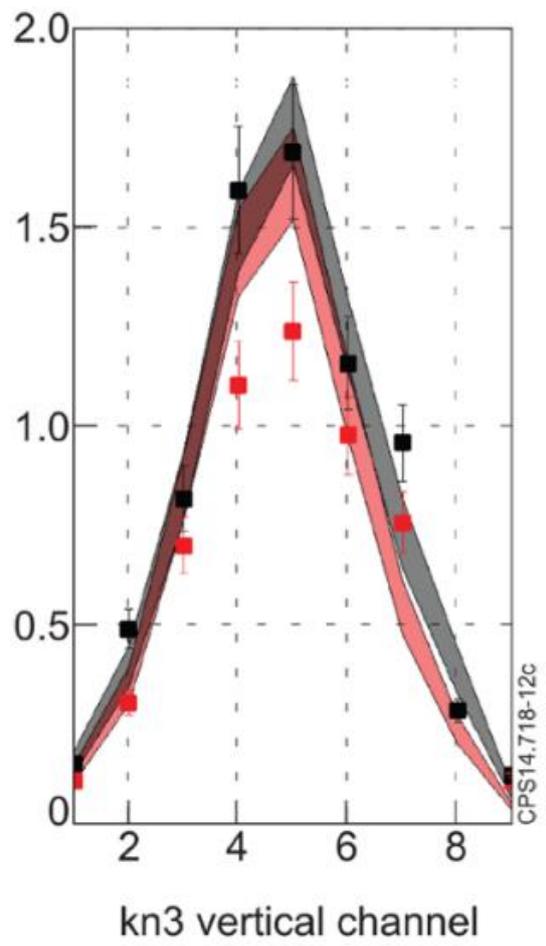

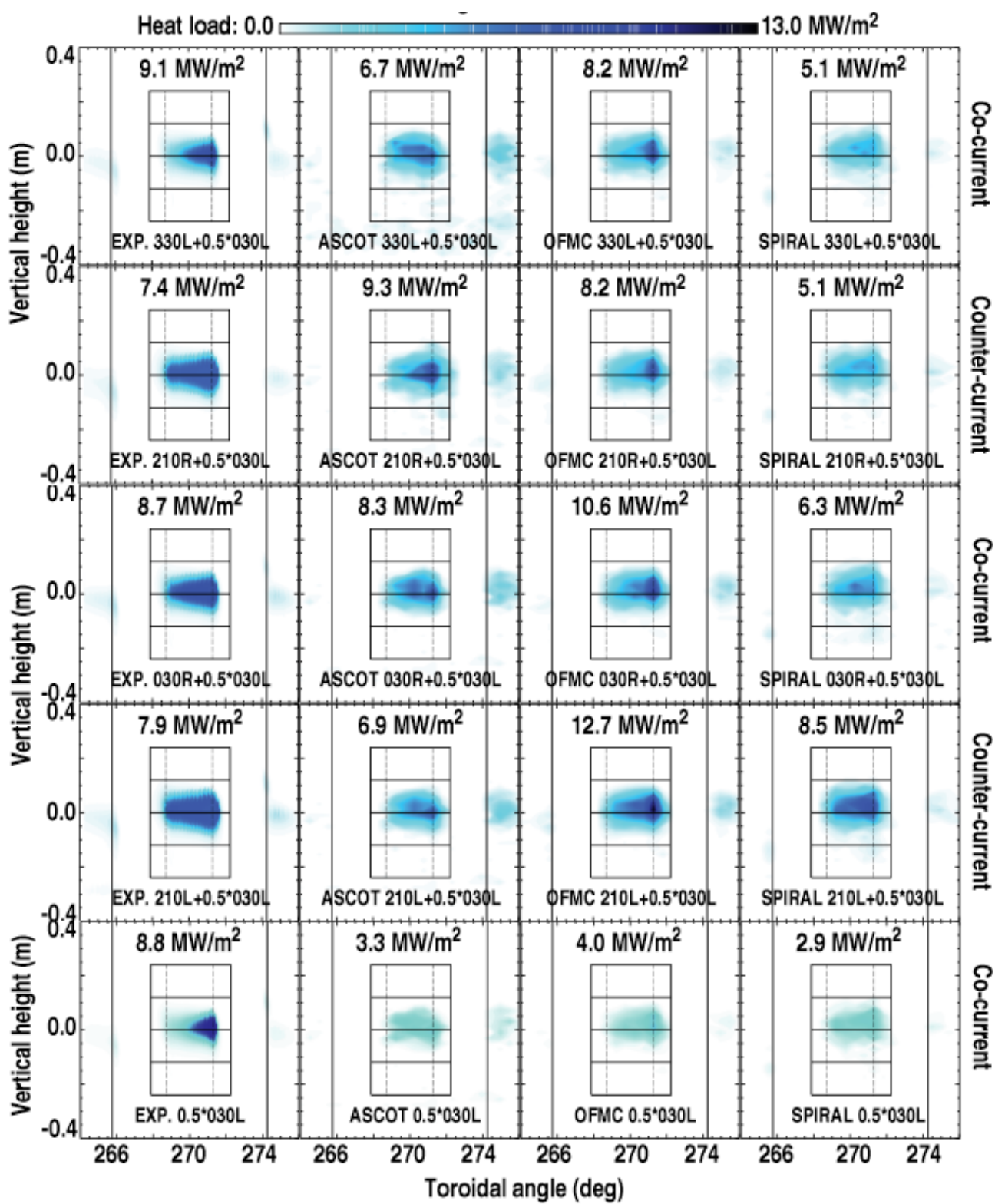

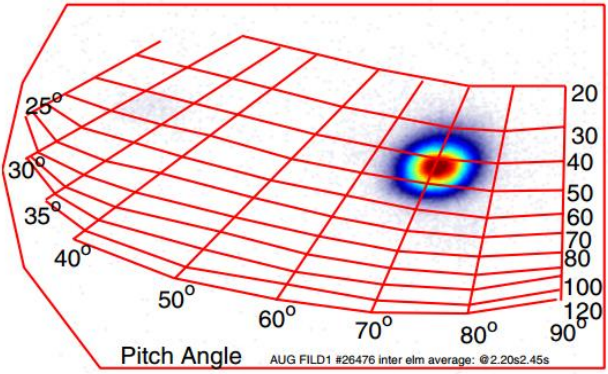 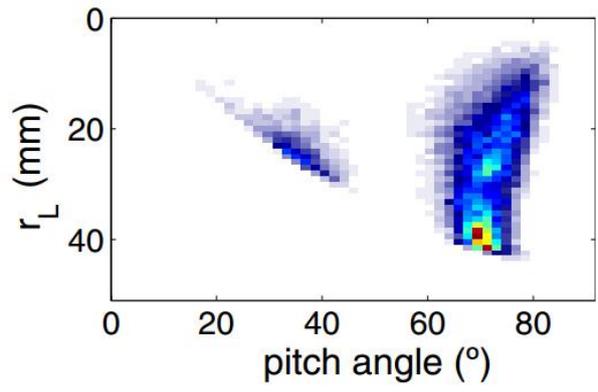

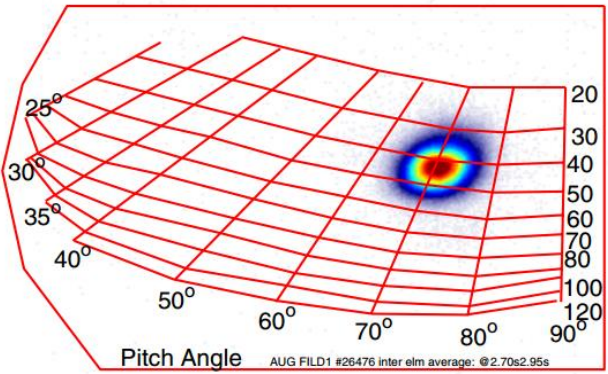 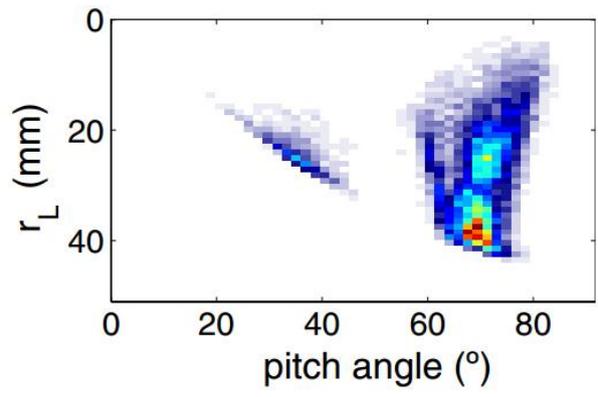

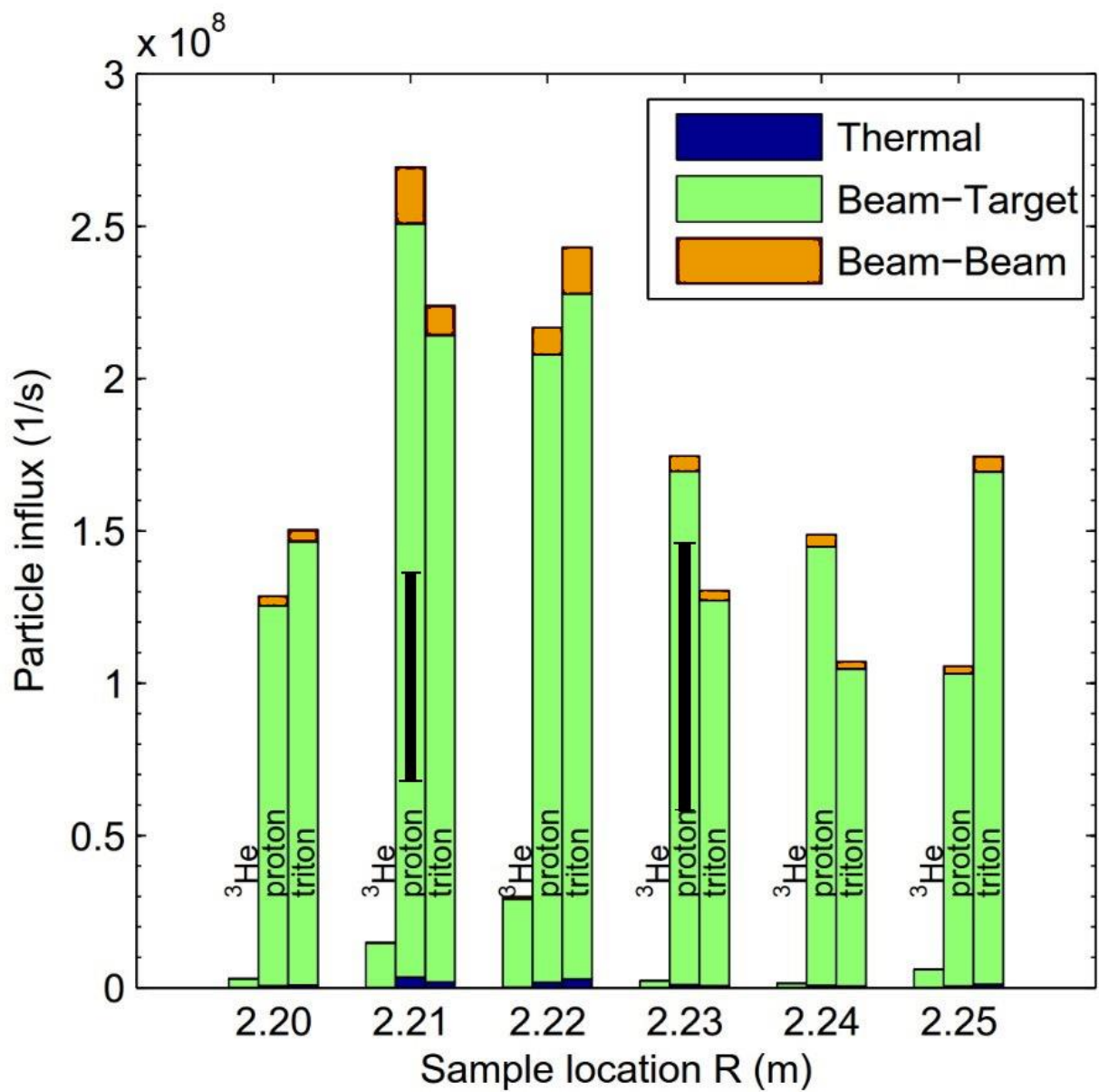

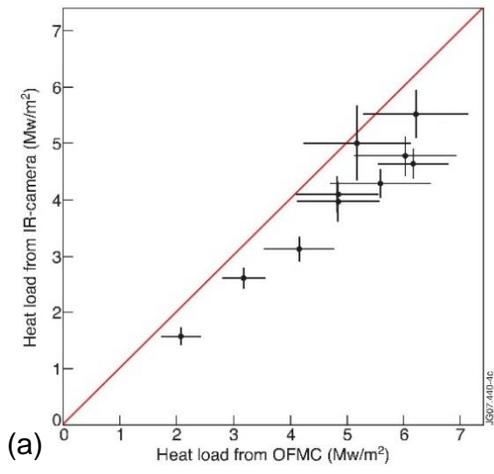 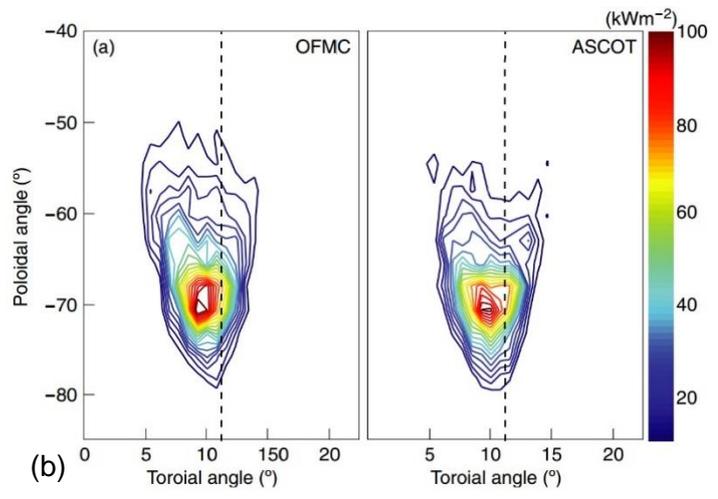

*Figure 1. From a suitable set of input data, the Monte Carlo orbit-following code ASCOT produces the distributions and history of the relevant ions and can record data for producing various synthetic diagnostics.*

*Figure 2. (a) Comparison of the measured and computed one-dimensional fast ion velocity distributions g(u) in ASDEX Upgrade discharge #24089. Two cases are shown, one where plasma is heated with two NBI sources (S3+S8, light/red) and another where only one source (S8, dark/blue) is applied. Solid line: TRANSP/NUBEAM; dashed line: ASCOT; bullets with error bars: CTS measurement; dotted line: bulk ions. The measured distribution is restricted to outside indicated bulk ion distributions. (b) Comparison of measured and synthetic CTS spectra for the same cases. Solid line: TRANSP/NUBEAM before consideration of the CTS data; dash-dotted line: TRANSP/NUBEAM after consideration of the CTS data consistent with the corresponding g(u) from (a), circles: measurement; dotted line: gyrotron frequency. Reproduced with permission from Ref. 4.*

*Figure 3: Comparisons of measured and simulated neutral fluxes in ASDEX Upgrade experiments. (a) Good quantitative agreement is seen in the case of relatively radial neutral beam and radial NPA sightline. (b) For a relatively tangential neutral beam, the simulations fail to capture the experimental features even qualitatively. In both cases, the simulations indicate a fairly uniform neutral flux spectrum up to the nominal beam energy, but as soon as either the sightline or the beam is turned away from the radial direction, the experimental signal drops, particularly at higher energies.*

*Figure 4. The signals measured by the KN3 neutron camera (squares) and the ASCOT-simulated synthetic signals (shaded areas) of the horizontal (left) and the vertical (right) neutron camera for two JET discharges. The error in the measured signals is assumed to be 10%, and the uncertainty in the simulated signals corresponds to the uncertainty in the plasma profile data. Reproduced with permission from Ref. 7.*

*Figure 5: Hot spots on the wall of DIII-D during the TBM mock-up experiments. First column: infrared measurements. Second column: ASCOT simulations of the corresponding cases. Third and fourth column: corresponding cases simulated by OFMC and SPIRAL, respectively. Different rows correspond to different beam geometries as indicated. Reproduced with permission from Ref. 8.*

*Figure 6. Comparison between experimental (left) and synthetic (right) FILD measurements for neutral beam Q5 in ASDEX Upgrade discharge #26476 without (top) and with (bottom) the magnetic perturbation. The fast ion flux is in arbitrary units in all the figures. In this context, the pitch angle is defined as $\xi = 180° − \arccos(v_\parallel/v)$. Reproduced with permission from Ref. 12.*

*Figure 7. Simulated and measured flux to the six samples inside the activation probe in ASDEX Upgrade. In each group of bars, the left one corresponds to $^3$He, the middle one to protons and the right one to tritons. The black bar is the measured proton flux (with uncertainty) in discharge #26229. Reproduced from Ref. 15.*

*Figure 8. (a) Comparison of the power load in JET calculated using the temperature rise on a poloidal limiter as measured by the IR camera and the peak power load as calculated with the OFMC code. The horizontal error bars are derived from the statistical noise in test particle*

*simulations and the vertical error bars include the errors in the IR temperature measurements and in relevant wall material parameters. Reproduced from Ref. 16. (b) Power load due to ripple-trapped beam ions, i.e., ions blocked between two adjacent field coils, as calculated by OFMC and ASCOT. The dashed line shows the toroidal location of ripple minimum, i.e., half-way between two toroidal coils. Reproduced with permission from Ref. 17.*